\documentclass[%
 reprint,
superscriptaddress,
 amsmath,amssymb,
 aps,
]{revtex4-2}

\usepackage{graphicx} % Required for inserting images
\usepackage{dcolumn}% Align table columns on decimal point
\usepackage{bm}% bold math
\usepackage{subfigure}
\usepackage{multirow}
\usepackage{amsmath,epsfig,amssymb,subfigure,bm,dsfont}
\usepackage{lipsum}

\begin{document}

\preprint{APS/123-QED}

\title{Quantum search in many-body interacting system with long-range interaction}

\author{Fan Xing}
\affiliation{ State Key Laboratory of Optoelectronic Materials and Technologies, School of Physics, Sun Yat-sen University, Guangzhou 510275, China}

\author{Yan Wei}
\affiliation{ State Key Laboratory of Optoelectronic Materials and Technologies, School of Physics, Sun Yat-sen University, Guangzhou 510275, China}

\author{Zeyang Liao}
\email{Corresponding author: liaozy7@mail.sysu.edu.cn}
\affiliation{ State Key Laboratory of Optoelectronic Materials and Technologies, School of Physics, Sun Yat-sen University, Guangzhou 510275, China}

\date{\today}

\begin{abstract}
Continuous-time quantum walks provide an alternative method for quantum search problems. Most of the earlier studies confirmed that quadratic speedup exists in some synthetic Hamiltonians, but whether there is quadratic speedup in real physical systems is elusive. Here, we investigate three physical systems with long-range atom-atom interaction which are possible good candidates for realizing the quantum search, including one-dimensional atom arrays either trapped in an optical lattice or coupled to waveguide near band edge or dispersively coupled to a good cavity. We find that all three systems can provide near-optimal quantum search if there is no dissipation. However, if the dissipation is considered only the latter two systems (i.e., waveguide-QED and cavity-QED systems) can still have high success probabilities because the latter two systems can significantly enhance the atom-atom interaction even if they are far apart and the spectra gap can be much larger which can reduce the search time and the effects of dissipation significantly. Our studies here can provide helpful instructions for realizing quantum search in real physical systems in the noisy intermediate-scale quantum era.
\end{abstract}

\maketitle

\section{Introduction}

Spatial search, the problem of finding a marked node in a graph, is one of the most widely used algorithms, which can be applied to search engines, combinatorial optimization (path navigation, recommendation systems), new materials and drug discovery, and many other mathematical problems (independent sets, SAT, dissimilar elements, subgroup finding, local search, weight determination), etc \cite{45nielsen2010,46knut1973}. For classical search algorithms, no shortcut is known and $O(n)$ queries are typically required where $n$ is the total number of elements. Quantum computation promises computational speedup over certain types of problems by leveraging quantum properties such as quantum superposition and entanglement \cite{45nielsen2010}. Grover first proposed a quantum search algorithm that requires only $O(\sqrt{n})$ oracle operations \cite{grover1996, 2Grover1997, 3Grover1998} and this algorithm can be realized based on quantum circuit model which attracts extensive interests \cite{Long2001, 38li2023, Huang2023}. In addition to solving search problems, the quantum search algorithm also provides new insight into ground state preparation \cite{Chowdhury2017,28lin2020}, high energy physics data processing \cite{Baker2021}, optimization problems \cite{30Somma2008,31lucas2014}, cryptography \cite{32zheng2023,33bijwe2020}, as well as solving NP-hard problems \cite{35Cerf2000, Kastella2011}. 

In addition to the quantum circuit model, the quantum computation can also be implemented via Hamiltonian evolution such as quantum adiabatic evolution \cite{Farhi2001,6Roland2002, Peng2008, 7Wubiao2016} and quantum random walk \cite{10Aharonov1993,11nayak2000,9kempe2003,portuga2018,portugal2013, PhysRevA.81.022308, PhysRevApplied.16.054036, PhysRevLett.128.050501}.  In 1998, Farhi and Gutmann proposed a quantum search algorithm based on continuous-time Hamiltonian evolution where they showed that quadratic speedup can also be achieved in a complete graph (i.e, every vertex has identical nonvanishing hopping rate to all other vertices) \cite{4FarhiGutmann1998}. It was then shown that for a hypercube graph where two vertices are connected with equal strength if and only if they differ in a single bit, the quantum spatial search algorithm based on Hamiltonian evolution can also provide a quadratic speedup \cite{Childs2002}. Based on a quantum analog of a discrete-time random walk (DTRW), Shenvi et al. also showed that optimal search can be achieved in a hypercube graph \cite{Shenvi2003}. In contrast to the DTRW algorithm, Childs and Goldstone proposed that spatial search can also be constructed via a continuous-time quantum walk (CTQW) where they showed that periodic lattice with dimension $d>4$ the marked node can be found in the optimal $O(\sqrt{n})$ times while $O(\sqrt{n}log^{\frac{3}{2}}n)$ running times are required for $d=4$ \cite{5ChildGoldstone2004}. However, when $d<4$ they showed that quadratic speedup is impossible. After that, many works have shown that spatial search by CTQW  is also optimal for other graph topologies such as fractal graph \cite{Agliari2010}, non-regular graph \cite{16novo2015}, Erd\"os-R\'enyi graph \cite{17Chakraborty2016}, and various network systems \cite{18Chakraborty2017,20Osada2020, 21Malmi2022}. In recent years, the necessary and sufficient conditions that a graph must fulfill for optimal quantum search have attracted intensive studies \cite{22Chakraborty2020, 23ambainis2020, 24Apers2022, PhysRevLett.114.110503, PhysRevLett.112.210502}.

In most of the previous studies, the quantum search algorithms are designed for pure mathematical models where the hopping rate is usually assumed to be a distance-independent constant and the dissipation of the system is usually ignored. In 2014, Childs and Ge showed that if the interaction strength decays as a quadratic power law with distance, the optimal spatial search can still be obtained in dimension $d=2$ system \cite{14Childs2014}. In 2021, Lewis et al. studied the spatial search on a closed one-dimensional (1D) spin chain with long-range interactions, where the system Hamiltonian $H_{0}=\sum_{i<j}J_{ij}(\vert i\rangle\langle j\vert+\vert j\rangle\langle i\vert)$ with $J_{ij}=|j-i|^{-\alpha}+|n-(j-i)|^{-\alpha}$, and they showed that when $\alpha<1.5$ the optimal spatial search exists but not for the case when $\alpha>1.5$ \cite{25Lewis2021}. Although this type of interaction may in principle be implemented in the linear ion trap chain system with a single-band M\o{}lmer-S\o{}rensen scheme \cite{Sorensen2000, Lewis2023}, its practical realization is challenging.

In this paper, we study the quantum search on three physical systems with different long-range interactions, and the aim is to find a physical realization as close as possible to the quantum search on the complete graph which is known to be optimal \cite{5ChildGoldstone2004}. The three physical systems we investigated include an atom chain either trapped in an optical lattice, coupled to a one-dimensional (1D) photonic waveguide near the photonic band edge, or dispersively coupled to a high-quality cavity. The results show that quadratic speedup with high success probability can be approached in all three systems if the dissipation is ignored. However, if the dissipation is considered, the success probability in the first system is extremely low, while the waveguide-QED and cavity-QED systems can still provide relatively high success probabilities due to the significant enhancement of collective long-range atom-atom interaction and the reduction of dissipation effects via vacuum engineering.  Our findings here can provide helpful instructions for the experimental realization of quantum spatial search in the noisy intermediate-scale quantum (NISQ) era \cite{Bharti2022}.

This article is organized as follows. In Sec. II we discuss the three different physical models and the quantum search algorithm based on CTQW. Then we numerically calculate the quantum search on the three different physical systems and show the condition for quadratic speedup over the classical search under the condition without dissipation noises in Sec. III and with dissipation noises in Sec. IV. Finally, we summarize our results.

\section{MODEL AND THEORY}

\subsection{Physical model}

Here, we consider the quantum search in a 1D atom array coupled to three photonic baths, i.e., free-space optical lattice, waveguide-QED, and cavity-QED systems, as shown in Fig. 1. We assume that all the atoms are identical with the same transition angular frequency $\omega_{a}$ and their nearest-neighbor separation is $d$. The atom-atom interaction can be long-range in all three systems, but how the interactions depend on the atom separation is quite different in these three systems. In a photonic environment, the electric field at position $\vec{r}$ generated by a dipole $\vec{\mu}_{1}$ at position $\vec{r}_{0}$ with oscillation frequency $\omega$ is given by $\vec{E}(\vec{r})=\omega^2\mu\vec{G}(\vec{r},\vec{r}_0,\omega)\cdot \vec{\mu}_{1}$ where $\mu$ is permeability of the photonic bath and $\vec{G}(\vec{r},\vec{r}_0,\omega_{a})$ is the dyadic Green function of the photonic environment \cite{Novotny_Hecht_2012}. When another dipole $\vec{\mu}_{2}$ with the same oscillation frequency is placed in position $\vec{r}$, the dipole-dipole coupling energy is then given by $\vec{\mu}_{2}\cdot \vec{E}(\vec{r})=\omega^2\mu \vec{\mu}_{2}\cdot\vec{G}(\vec{r},\vec{r}_0,\omega)\cdot \vec{\mu}_{1}$.  Thus, the general expression for the effective atom-atom interaction in arbitrary photonic bath is then given by  \cite{Sheremet2023}
\begin{equation}
V_{ij}=J_{ij}+i\Gamma_{ij}/2=(\omega_{a}^{2}\mu/\hbar)\vec{\mu}^{*}_{i}\cdot \vec{G}(\vec{r}_{i},\vec{r}_j,\omega_{a})\cdot \vec{\mu}_{j}
\end{equation}
where the real part $J_{ij}$ is the coherent dipole-dipole interaction and the imaginary part $\Gamma_{ij}$ is the incoherent collective dissipation rate mediated by the photonic baths. $\vec{\mu}^{*}_{i}$ is the transition dipole moment of the $i$th atom at position $\vec{r}_{i}$ which is assumed to be perpendicular to the atom arrays and $\hbar$ is the reduced Planck constant. The effective atom-atom interactions and the collective decays for the three cases are shown in Table I. Before proceeding, we first introduce the definition of the so-called ``long-range'' interaction used in current work. Here, ``long-range'' is used to denote generic non-local couplings, i.e., beyond on-site or nearest-neighbor couplings \cite{RevModPhys.95.035002}. 

\begin{figure}[htbp]
\label{Fig1} \centering\includegraphics[width=8.6cm]{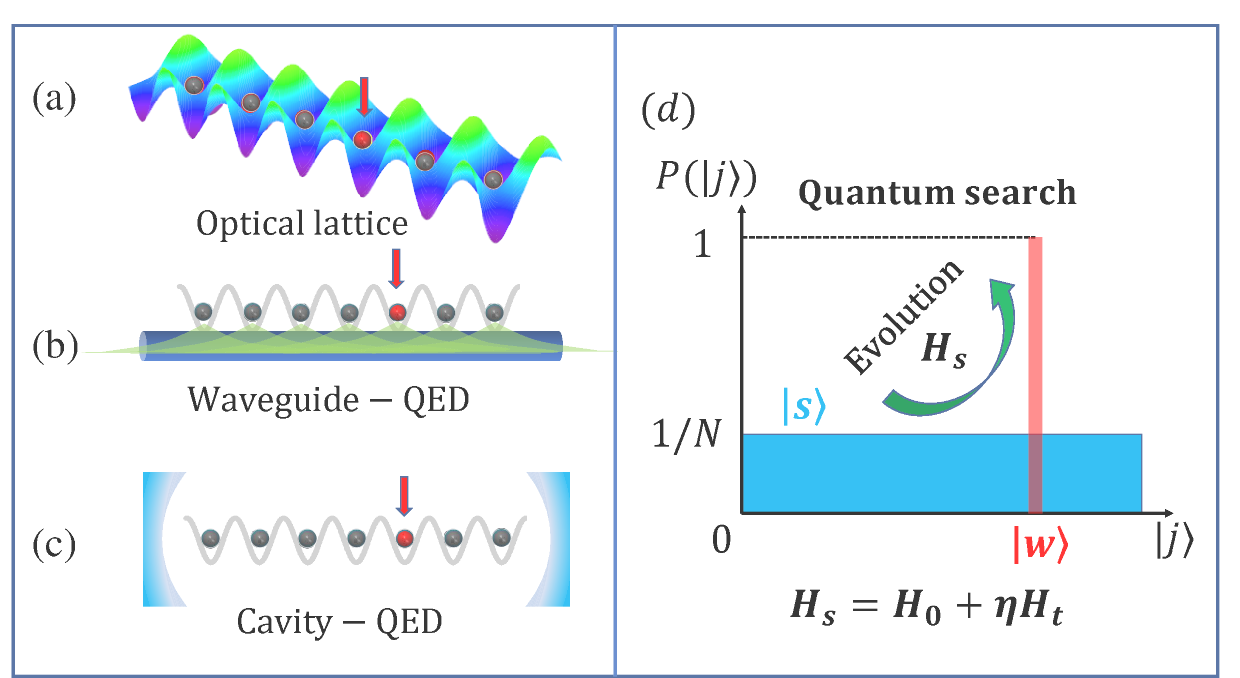}
\caption{(Color online)  Quantum search in three different physical systems: (a) 1D cold atom array trapped in an optical lattice with a mixture of power-law interaction between atoms; (b) 1D atom array coupled to a waveguide with an exponentially decaying interaction between atoms; (c) 1D atom array dispersively coupled to a cavity with ideal long-range interactions between atoms. (d) Schematic diagram of the quantum search algorithm where the system evolves from an equal superposition state $|s\rangle$ to the target state $|w\rangle$ under the evolution of search Hamiltonian $H_s$.}
\label{1}
\end{figure}

\begin{table*}
\renewcommand{\arraystretch}{2.0}
    \centering
   \begin{tabular}{|c|c|c|}
   \hline 
   Physical System   &  Coherent interaction  & Collective dissipation \\
\hline 
Free-space optical lattice & $J_{ij}=\frac{3\gamma}{4}\big [-\frac{\cos(k_{a}r_{ij})}{k_{a}r_{ij}}+\frac{\sin(k_{a}r_{ij})}{(k_{a}r_{ij})^2}+\frac{\cos(k_{a}r_{ij})}{(k_{a}r_{ij})^3}\big ]$  & $\Gamma_{ij}=\frac{3\gamma}{2}\big [\frac{\sin(k_{a}r_{ij})}{k_{a}r_{ij}}+\frac{\cos(k_{a}r_{ij})}{(k_{a}r_{ij})^2}-\frac{\sin(k_{a}r_{ij})}{(k_{a}r_{ij})^3}\big ]$  \\ 
\hline 
Waveguide-QED above bandgap & $J_{ij}=\frac{\Gamma}{2}\sin(k_{a} r_{ij})$  & $\Gamma_{ij}\rightarrow \frac{\Gamma}{2}\cos(k_{a} r_{ij})$  \\ 
\hline 
Waveguide-QED within bandgap & $J_{ij}=\frac{\Gamma}{2}e^{-\kappa r_{ij}}$  & $\Gamma_{ij}\rightarrow 0$  \\ 
\hline 
Cavity-QED with dispersive coupling & $J_{ij}=\frac{g^2}{\Delta}$  & $\Gamma_{ij}\approx\frac{3\gamma}{2}\big [\frac{\sin(k_{a}r_{ij})}{k_{a}r_{ij}}+\frac{\cos(k_{a}r_{ij})}{(k_{a}r_{ij})^2}-\frac{\sin(k_{a}r_{ij})}{(k_{a}r_{ij})^3}\big ]$  \\ 
\hline
\end{tabular}
    \caption{Coherent atom-atom interaction strengths and the incoherent dissipation for different physical systems with long-range interaction.}
    \label{tab:my_label}
\end{table*}

For an atom array trapped by the 1D optical lattice, the atom-atom interaction is a mixture of power-law interaction, i.e., $J_{ij}=\frac{3\gamma}{4}\big [-\frac{\cos(k_{a}r_{ij})}{k_{a}r_{ij}}+\frac{\sin(k_{a}r_{ij})}{(k_{a}r_{ij})^2}+\frac{\cos(k_{a}r_{ij})}{(k_{a}r_{ij})^3}\big ]$ and the collective decay $\Gamma_{ij}=\frac{3\gamma}{2}\big [\frac{\sin(k_{a}r_{ij})}{k_{a}r_{ij}}+\frac{\cos(k_{a}r_{ij})}{(k_{a}r_{ij})^2}-\frac{\sin(k_{a}r_{ij})}{(k_{a}r_{ij})^3}\big ]$  where $\gamma$ is the spontaneous decay rate of a single atom in the free space, $k_a=\omega_{a}/c$ (c is the speed of light) and $r_{ij}=|\vec{r}_{i}-\vec{r}_{j}|$ \cite{Ficek2004,Liao2014}. When $r_{ij}\gg \lambda_{a}$ ($\lambda_{a}$ is the wavelength related to $\omega_{a}$), $J_{ij}\propto 1/r_{ij}$, $\Gamma_{ii}\rightarrow \gamma$ and $\Gamma_{ij}\rightarrow 0$ when $i\neq j$. For the power-law interaction decays as $1/x^{\alpha}$, it can be called strong long-range interaction when $\alpha<d$, and weak long-range interaction when $\alpha>d$ where $d$ is the dimension of the system according to Ref. \cite{RevModPhys.95.035002}.

For the atom-waveguide coupling system, the effective atom-atom interaction in the case when the atomic transition frequency is above the cutoff frequency of the waveguide is given by $V_{ij}=(\Gamma/2)e^{ik_{a}r_{ij}}$ where $\Gamma$ is the decay rate into the waveguide mode \cite{41Liao2015, Caneva2015, 42Liao2016, Xing2022, Xing2024}. The coherent part of the interaction is also long-range but it is periodically varying with atom distance. If the atomic transition frequency is slightly below the cutoff frequency $\omega_{c}$ of the waveguide, the atom is mainly coupled to the modes around the band edge due to the van Hove singularity of the density of states and the radiation dissipation can be almost completely suppressed \cite{Hulet1985, Yablonovitch1987}. In this case, the effective atom-atom interaction is given by $V_{ij}=(\Gamma/2)e^{-\kappa r_{ij}}$ which is exponentially decaying as the atom separation increases with decaying factor $\kappa=\sqrt{\omega_{c}^2-\omega_{a}^{2}}/c$ \cite{Shahmoon2013,39douglas2015}. Although it decreases exponentially with atom separation, the effective coupling length depends on the detuning $\Delta=\omega_c-\omega_{a}$ which can still become effective long-range if the detuning is very small. For example, if $\kappa$ is very small, there is still a significant interaction between the first and the last atoms which is a long-range interaction according to the definition of long-range used in current work. The advantage of the case, when the atomic transition is below the cutoff frequency, is that the spontaneous decay can be almost completely inhibited.

For the atom-cavity coupling system, we consider the case that multiple two-level atoms dispersively couple to a common cavity mode which can produce effective infinite long-range interaction between atoms $J_{ij}^{C}=g^2/\Delta$, where $g$ is the atom-cavity coupling strength and $\Delta$ is the detuning between the atomic transition frequency and the cavity frequency \cite{agarwal_2012}. It is clearly seen that the atom-atom interaction in this case is effectively infinite long-range which does not decrease with atom distance. Since the atomic transition frequency is significantly detuned from the cavity mode, the dissipation to the cavity mode is largely suppressed but the dissipation to other directions can still occur and therefore the dissipation is about the same as that in the free space which is shown in Table I \cite{Heinzen1987}.

\subsection{Quantum search based on CTQW}

Considering that there are n atoms with a single excitation in the above systems, it can be mapped to a graph, $G$, with $n$ vertices each representing a single excited atom. The basis state $\vert j \rangle=\vert 0...1_{j}...0 \rangle$ with $j=1,2,\cdots,n$ denoting that the $j$th atom is in the excited state while other atoms are in the ground state. According to the continuous-time quantum walks spatial search algorithm proposed by Childs and Goldstone \cite{5ChildGoldstone2004}, we can construct the search Hamiltonian
\begin{equation}
H_{s}=H_{0}+\eta H_{t}.
\label{Hs}
\end{equation}%
Here, $H_{0}=\sum_{ij}V_{ij}\sigma_{eg}^{i}\sigma_{ge}^{j}$ is the effective system (graph) Hamiltonian in the interaction picture and $V_{ij}$ is the effective coupling strength given by Eq. (1) and the detail expressions of $V_{ij}$ for the three different physical systems are shown in Table I. $\sigma_{eg}^{i}=\vert e\rangle_{i}\langle g\vert$ and $\sigma_{ge}^{j}=\vert g\rangle_{j}\langle e\vert$, so $H_{0}$ can be written as $H_{0}=\sum_{ij}V_{ij}\vert i\rangle\langle j\vert$. $H_{t}=\vert w \rangle \langle w \vert$ is the target state Hamiltonian with $w\in\{1,...n\}$ denoting that the $w$th atom is excited and other atoms are in the ground state, and parameter $\eta$ denoting the relative strength of the two Hamiltonian. In practical realization, the target Hamiltonian is realized by shifting the transition frequency of the $w$th atom by applying an external control field. Here, we should mention the connection between our search Hamiltonian shown in Eq. (2) with the search Hamiltonian used in the previous research, i.e., $H_s=\beta H_{0}+H_{t}$ where $H_{0}=\sum_{ij}\vert i\rangle\langle j\vert$ without dissipation \cite{5ChildGoldstone2004}. In the case of the complete graph without dissipation, i.e, $J_{ij}$ is a constant (e.g., $J_{ij}=J_0$), it is readily seen that when $\eta=J_{0}/\beta$, both Hamiltonian are actually equivalent and only differ by a constant rescaling factor.

The typical initial state of the system is an equal superposition of one-excitation state i.e., $\vert s \rangle = \frac{1}{\sqrt{n}} \sum_{j=1}^{n} \vert j \rangle$. For a quantum search algorithm, we require that the system can evolve to the target state $|w\rangle$ with high fidelity at certain time under the Hamiltonian shown in Eq. (2). If the atom-atom interaction is ideally long-range with $J_{ij}$ being a positive constant, it is not difficult to prove that $\vert s \rangle$ is the eigenstate of the coherent part of $H_0$ with the largest eigenvalue. When $\eta\rightarrow \infty$, $|w\rangle$ is the eigenstate of $H_s$ with the largest eigenvalue. At certain finite values of $\eta$, neither $|s\rangle$ nor $|w\rangle$ states but their superposition is the eigenstate of $H_s$. We need to find the optimal value of $\eta$ such that the relevant eigenstates have maximum overlap with both the $|s\rangle$ and $|w\rangle$ states. Under this condition, the system can oscillate between the $|s\rangle$ and $|w\rangle$ states and at certain time the system can evolve from the initial state $|s\rangle$ to the target state $|w\rangle$ with high fidelity.

The goal of the quantum search is obtaining the target state $|w\rangle$ from the initial state $|s\rangle$ with maximum fidelity, and the searching procedure is as follows \cite{5ChildGoldstone2004}.
(1)	Choose optimal $\eta$: Select an appropriate range of $\eta$ to calculate the eigenvalues $E_{j}$ and eigenstates $|\phi_{j}\rangle$ of the coherent part of $H_{s}$. Here we mainly care about the eigenstates $|\phi_0\rangle$ and $|\phi_1\rangle$ with the largest and the second largest eigenvalues, i.e., $E_0$ and $E_1$. If $\eta$ varies from zero to infinity, $|\phi_0\rangle$ switch from $|s\rangle$ to $|w\rangle$ and at a certain value of $\eta$ the two eigenstates $|\phi_0\rangle$ and $|\phi_1\rangle$ cross over and these two eigenstates can both have substantial overlap on both $|s\rangle$ and $|w\rangle$. The crossover occurs under the condition that $\Delta E=E_{0}-E_{1}$ is the minimum from which we can determine the optimal parameter $\eta_{opt}$. Under this Hamiltonian, the system can evolve from the state $\vert s \rangle$ to the target state $\vert w \rangle$ with high fidelity at a time proportional to the inverse of the energy gap $\Delta T_{min}=\pi/\Delta E_{min}$.
(2)	To prove that the system can indeed evolve to the target state with high fidelity, we let the system evolve under the search Hamiltonian $H_{s}= H_{0}+\eta_{opt} H_{t}$ from the initial state $\vert s \rangle$ for a time $t$, obtaining the final state $|\psi_{f}\rangle=e^{-iH_{s}t} \vert s \rangle$.
(3)	Make a projection measurement and the success probability of finding the target state $\vert w \rangle$ is $F=|\langle w | e^{-iH_{s}t} |s \rangle|^2$. Define the time corresponding to the maximum fidelity as the optimal search time $T_{opt}=t_{F_{max}}$ and compare this search time with $\Delta T_{min}$.
(4) Vary the system size $n$ and repeat the above procedures to determine the searching time as a function of $n$. Finally, we fit the curves and see whether there is a quadratic speedup or not.

\section{Quantum search without dissipation-numerical simulation}

In this section, we numerically study the optimal searching time for three different interacting systems shown in Fig. 1.  We assume that in all three cases, atom separation is $d=\lambda_{a}$ which is the wavelength corresponding to the atomic transition frequency $\omega_{a}$. In the following calculations, we let $\lambda_{a}=1$ for simplicity. Thus the distance between the $i$th and $j$th atoms is given by $r_{ij}=|i-j|$. We vary the atom number from 100 to 1000 and calculate the search time for the three interaction systems using both  $\Delta T_{min}$ and $T_{opt}$, as well as the maximum fidelity of finding the target state. To compare our results with the previous research for the ideal complete graph \cite{5ChildGoldstone2004}, we first ignore the dissipation of the system (i.e. $H_{0}=\sum_{ij}J_{ij}\vert i\rangle\langle j\vert$) in this section and the effects of dissipation will be considered in the next section.

\begin{figure*}[htbp]
\label{Fig2} \centering\includegraphics[width=17cm]{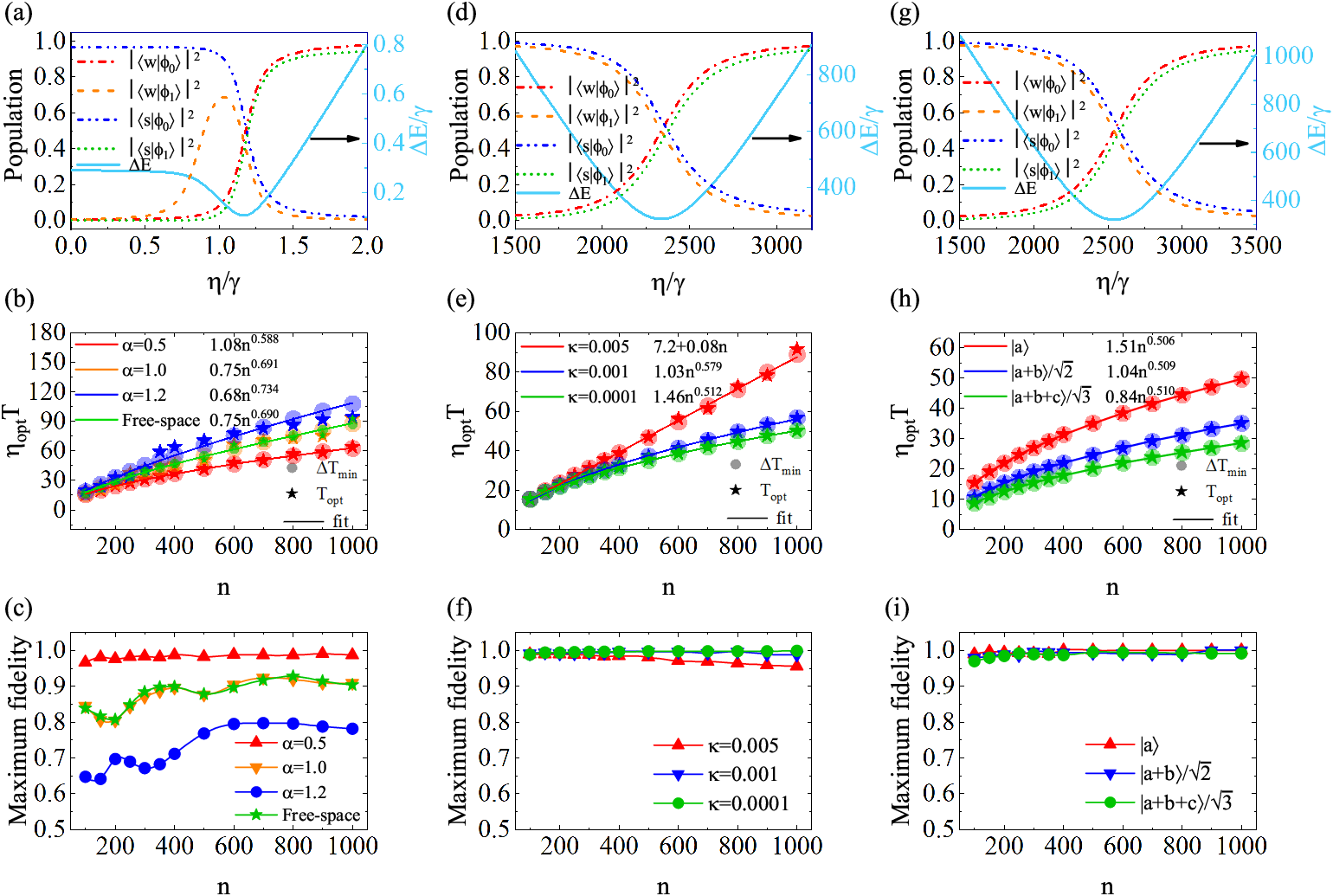}
\caption{(Color online)  (a, d, g) Determining the optimal $\eta_{opt}$ for the three different physical systems: (a) optical lattice, (d) waveguide-QED with $\kappa=0.001$, and (g) cavity-QED. $n=256$ for all three cases.  Optimal searching time and maximum fidelity for: (b, c) mixture-power-low interaction in free space optical lattice and pure-power-law interaction for $\alpha=0.5, 1.0, 1.2$; (e, f) exponentially decaying interaction in the waveguide-QED system;  (h, i) infinite long-range interaction in the cavity-QED system.}
\label{2}
\end{figure*}

\subsection{Atom array in optical lattice: multi-power-law interaction}

The spatial search on a closed one-dimensional (1D) spin chain whose Hamiltonian $H_{0}=\sum_{i<j}J_{ij}(\vert i\rangle\langle j\vert+\vert j\rangle\langle i\vert)$ with $J_{ij}=|j-i|^{-\alpha}+|n-(j-i)|^{-\alpha}$ has been studied \cite{25Lewis2021}. Here, we discuss a slightly different power-law interaction which may be much easier to implement. As is known to us, an atom array in the free space or trapped by an optical lattice can have dipole-dipole interaction induced by the free-space vacuum which is a mixture of power-law as shown in Table I \cite{Ficek2004, Liao2014}. When $k_{a}r_{ij}\gg 1$, $J_{ij}\propto r_{ij}^{-1}$, while when $k_{a}r_{ij}\ll 1$, $J_{ij}\propto r_{ij}^{-3}$. Since we mainly consider the case when the nearest-neighbor atom separation is $\lambda_a$, the atom-atom interaction $J_{ij}$ is dominated by the first term proportional to $r_{ij}^{-1}$. Different from the symmetric Hamiltonian used in Ref. \cite{25Lewis2021}, there is not known analytical solution for the multi-power-law interaction Hamiltonian and here we resort to the numerical simulations.

We first calculate the eigenvalues and eigenvectors of the search Hamiltonian shown in Eq. (2) when $\eta$ varies from 0 to $\infty$. Taking $n=256$ and $w=20$ as an example, the energy difference $\Delta E$ between the largest and second largest eigenvalues as a function of $\eta$ is shown in Fig. 2(a) from which we can see that there is a minimum energy gap. The wavefunction overlaps between the two eigenstates ($|\phi_0\rangle$ and $|\phi_1\rangle$) and the initial state $|s\rangle$ and target state $|w\rangle$ are also shown in Fig. 2(a). It is clearly seen that when $\Delta E$ is minimum, the two largest eigenstates have almost equal overlaps with $|s\rangle$ and $|w\rangle$ states which is the requirement for optimal quantum search with high success probability. From this, we can determine that the optimal $\eta_{opt}\approx 1.17\gamma$ when $n=256$ and $|w\rangle=|20\rangle$. Now, with different sizes of atom array, we first determine the optimal $\eta_{opt}$ for each $n$ and then numerically calculate the optimal searching time $T_{opt}$ by evolving the system under the search Hamiltonian shown in Eq. (2). The results are shown as the green solid stars in Fig. 2(b). To compare the search time on an equal footing, we have multiplied the searching time by $\eta_{opt}$ to eliminate the effect of $\eta_{opt}$. In addition to calculating the search time by Hamiltonian evolution, we also evaluate the search time by simply calculating $\Delta T_{min}$ from the minimum energy gap, and the results are shown as the green dots in Fig. 2(b). It is clearly seen that the search time calculated by these two methods coincides with each other very well. By fitting the dots (green solid line), we can find that $\eta_{opt}t_{opt}=0.75\times n^{0.690}$ which is slightly slower than the quadratic acceleration of Grover's search algorithm but is faster than the classical search algorithm. For comparison, we also show the results of pure power-law interaction (i.e., $J_{ij}\propto r_{ij}^{-\alpha}$) with $\alpha=0.5, 1.0$, and $1.2$. We can see that as $\alpha$ decreases, the search time also decreases and approaches the quadratic acceleration. The result when $\alpha=1$ is very close to our free-space case which confirms that the first term in $J_{ij}$ dominates when $d=\lambda_{a}$.

We also calculate the maximum fidelity of the evolving final state with the target state and the results are shown in Fig. 2(c). For the multi-power-law interaction in the free-space case, the fidelity is between 80\% and 90\% when the system size varies from 100 to 1000. This fidelity is high enough for practical realization. In the pure power-law case, when $\alpha$ decreases, the success probability of finding the target node increases. Particularly, when $\alpha=0.5$, the fidelity is close to the unit. Again, the success probability when $\alpha=1.0$ is almost the same as that in our case.

\subsection{Waveguide-QED: exponentially decaying interaction}

It is known that the atom-atom interaction in the waveguide-QED system can also be long-range. When the atom transition frequency is above the cutoff frequency of the single-mode optical waveguide, the coherent atom-atom interaction is given by $J_{ij}=(\Gamma/2)\sin(k_a r_{ij})$ and the collective dissipation $\Gamma_{ij}=(\Gamma/2)\cos(k_a r_{ij})$  which are periodic functions of the atom separation (see Table I) \cite{41Liao2015, Caneva2015, 42Liao2016, Xing2022, Xing2024}. We find that under this Hamiltonian, the equal superposition state $|s\rangle$ is not an eigenstate of $H_0$ and the optimal quantum search can not be found. 

Instead, we consider the case when the atomic transition frequency is slightly less than the cutoff frequency of the single-mode optical waveguide where the atom-atom interaction exponentially decays with the atom separation, i.e.,  $J_{ij}=(\Gamma/2)e^{-\kappa r_{ij}}$ with vanishing dissipation which is presented in the third row of Table I \cite{Shahmoon2013,39douglas2015}.  Here, in the numerical simulations, we choose $\Gamma=20\gamma$ and three different values of $\kappa$ (i.e., $\kappa=0.005, 0.001$, and $0.0001$). Here we choose these three different values of $\kappa$ to demonstrate the search time transiting from the linear function of $n$ to the approximate quadric function of $n$. The parameter $\kappa$ depends on the transition frequency of the atom and the cutoff frequency of the waveguide, i.e., $\kappa=\sqrt{\omega_{c}^{2}-\omega_{a}^{2}}/c\approx\sqrt{2\omega_{a}\Delta\omega_{ca}}/c$ where $\Delta\omega_{ca}=\omega_{c}-\omega_{a}$ is the energy difference between the cutoff frequency and the atom transition frequency \cite{Shahmoon2013,39douglas2015}. In practical realization, we can control the detuning $\Delta\omega_{ca}$ to tune the value of $\kappa$. 
It should be noted that we set $\lambda_a=1$ in the current work and therefore $\kappa=0.001$ is actually in units of $1/\lambda_{a}$. Then, we have $\Delta\omega_{ca}\approx\kappa^{2}\omega_{a}/2$. Supposing that $\lambda_a=1\mu m$, the atomic transition frequency $\omega_a=6\pi\times10^{14} Hz$. To obtain $\kappa=0.001$,  it requires that the energy difference $\Delta\omega_{ca}\approx3\pi\times10^{8}Hz$. The energy difference $\Delta\omega\approx75\pi\times10^{8}Hz$ results in $\kappa=0.005$ and $\Delta\omega\approx3\pi\times10^{6}Hz$ yields $\kappa=0.0001$. 

In Fig. 2(d), we show the energy difference $\Delta E$ between the largest and second largest eigenvalues of the search Hamiltonian as a function of $\eta$ when $n=256$ and $\kappa=0.001$. From the figure, we can see that there is a minimum energy gap $\Delta E$ occurring at $\eta_{opt}\approx 2335\gamma$ where the corresponding eigenstates have significant overlaps with the initial state $|s\rangle$ and the target state $|w\rangle$ ($|w\rangle=|20\rangle$ in our numerical calculations). It is noted that the optimal value of $\eta$ here is much larger than that used in the free-space optical lattice with the same number of atoms. There are two main reasons why $\eta_{opt}$ is much larger here. First, due to the confinement of the photon field, the interaction strength between the atom and the photon can be much larger which leads to a larger $\Gamma$. Second, when $\kappa$ is small, the atom-atom interaction decays slower than that in the free space which results in a much stronger atom-atom interaction even if two atoms are far apart. Due to these two factors, the energy scale of the system is much larger which requires a much larger $\eta_{opt}$ for observing the cross-over behavior, and the minimum energy gap $\Delta E$ in this case is also three orders of magnitude larger than that in the free-space optical lattice. Actually, the large value of $\eta_{opt}$ can also be predicted from the previous studies for the complete graph because the system here can be well described by a complete graph When $\kappa\rightarrow 0$. It has been shown in the complete graph that the optimal search is reached when $\beta\approx 1/n$ \cite{5ChildGoldstone2004} which corresponds to $\eta_{opt}\approx n\Gamma/2$ in our case. Thus, the optimal value of $\eta$ depends both on the number of atoms and the atom-atom interaction strength $\Gamma$. Here, we choose $\Gamma=20\gamma$ and thus $\eta_{opt}\approx 10n\gamma$. Indeed, for the case with $n=256$, when $\kappa$ is reduced from $0.001$ to $0.0001$, $\eta_{opt}$ increases from $2335\gamma$ to $2520\gamma$ which is approaching the value of $2560\gamma$ in the complete graph.

After determining the optimal values of $\eta$ for each $n$, we numerically evolve the system under the search Hamiltonian shown in Eq. (2) with three different values of $\kappa$. We can determine the optimal search time as a function of the number of atoms $n$ as shown in Fig. 2(e). Again, to compare the search time on an equal footing, here we also multiply the search time with $\eta_{opt}$. By fitting the curves, we find that when $\kappa=0.005$, the search time $\eta_{opt}t_{opt}\propto 0.08n$ is a linear function of number of sites. Thus, there is no quadratic speedup over the classical search ($\eta_{opt}t\propto 0.5n$) but the slope in our quantum search here is about 6 times smaller which indicates the speed up of the search. However, when we decrease the value of $\kappa$, we find that the search time also decreases. When $\kappa=0.001$, the search time $\eta_{opt}t_{opt}\approx 1.03n^{0.579}$ and when $\kappa=0.0001$, the search time $\eta_{opt}t_{opt}\approx 1.46n^{0.512}$ which is apparently approaching to the quadratic speed up and also very close to the optimal quantum search time predicted in the complete graph ($t_{opt}\approx (\pi/2)n^{1/2}$) \cite{5ChildGoldstone2004}. The maximum fidelity (i.e., the success probability) as a function of $n$ is shown in Fig. 2(f) from which we can see that the fidelities for all three cases are very close to unit. Thus, quantum speedup is also possible if the atom array is coupled to the waveguide near the photonic band edge.

\subsection{Cavity-QED: infinite long-range interaction}

For an atom chain dispersively coupled to a cavity, the atom-atom interaction induced by the common cavity mode is ideally infinite long-range with $J_{ij}=g^2/\Delta$ which is a constant for arbitrary pair of atoms \cite{agarwal_2012}. Thus, this system is a possible physical realization of a quantum search in a complete graph. Here in the numerical simulation, we assume that $J_{ij}=10\gamma$ which is experimentally achievable \cite{44nature2019}.  To determine the optimal parameter $\eta$, we also calculate the energy gap $\Delta E$ between the largest and second largest eigenvalues of the search Hamiltonian when $\eta$ varies from 0 to $\infty$. Again, we find that there is also a minimum energy gap when $\eta_{opt}=2540\gamma$ (for $n=256$, $\vert w\rangle=\vert 20\rangle$) (Fig. 2(g)). It is readily seen that $\eta_{opt}$ here is also very close to the value ($2560\gamma$) as predicted by the precious studies for the complete graph \cite{5ChildGoldstone2004}. Similar to the waveguide-QED system, $\eta_{opt}$ here is also much larger than that in the free space system due to the much larger long-range atom-atom interactions.  We also find that the minimum energy gap is three orders of magnitude larger than that in the free space case which can significantly reduce the search time. Then we numerically evolve the system and calculate the optimal searching time $\Delta T_{min}$ and $T_{opt}$  for three different target states and the results are shown in Fig. 2(h). The three target states include searching for a single node $|w\rangle=|a\rangle$, superposition of two nodes  $|w\rangle=\frac{1}{\sqrt{2}}(|a\rangle+|b\rangle)$ and three nodes $|w\rangle=\frac{1}{\sqrt{3}}(|a\rangle+|b\rangle+|c\rangle)$ with $a=20$, $b=40$ and $c=60$ as an example. From the figure, we can see that $\Delta T_{min}$ and $T_{opt}$ coincide with each other and there is a clearly quadratic speedup for searching all three different target states. For a single marked node, by fitting the curve we can obtain  $\eta_{opt}t_{opt}\approx 1.51 \times n^{0.506}$ which is very close to the theoretical predicted behavior ($t_{opt}\approx (\pi/2)n^{1/2}$) in the complete graph \cite{5ChildGoldstone2004}. We also see that $\eta_{opt}t_{opt}=1.04 \times n^{0.509}$ and  $\eta_{opt}t_{opt}=0.84 \times n^{0.510}$ for the cases with two and three marked nodes, respectively. It turns out that searching a superposition state is even faster than searching a single node because this kind of target state has a larger overlap with the initial state. For searching multiple nodes in a complete graph, it has been shown that the searching time is given by $t=(\pi/2)\sqrt{n/k}$ where $k$ vertices are marked \cite{wong2016}. According to this prediction, the search time is reduced by $1/\sqrt{k}$ when $k$ nodes are marked compared with only one node is marked. Indeed, our numerical results are consistent with this prediction. The optimal search time for one node here is about $1.51 \times n^{0.506}$. Accordingly, the optimal searching times for $k=2$ and $k=3$ are predicted to be about $1.06n^{0.506}$ and $0.86n^{0.506}$ which are very close to our numerical results $1.04n^{0.509}$ and $0.84n^{0.510}$, respectively.
The maximum fidelities (or success probabilities) for finding the target states for all three cases are close to 100\% as shown in Fig. 2(i).

In both the waveguide-QED with $\kappa\rightarrow 0$ and the cavity-QED systems with infinite long-range interaction, $\eta_{opt}$ is required to be much larger than $\gamma$ which may pose a great challenge for the experimental realization of the quantum search algorithm in these two models. However, this requirement is not impossible. For example, in \cite{PhysRevApplied.18.034057}, researchers have demonstrated the tuning of transition frequency by more than 40 MHz in the superconducting qubits while its relaxation time is about 10 KHz. Another possible candidate is the Rydberg atom whose Stark effect can be of the order of GHz while its decay rate can be as low as KHz \cite{Jarisch2018, evered2023}.

\begin{figure}[htbp]
\label{Fig4} \centering\includegraphics[width=\columnwidth]{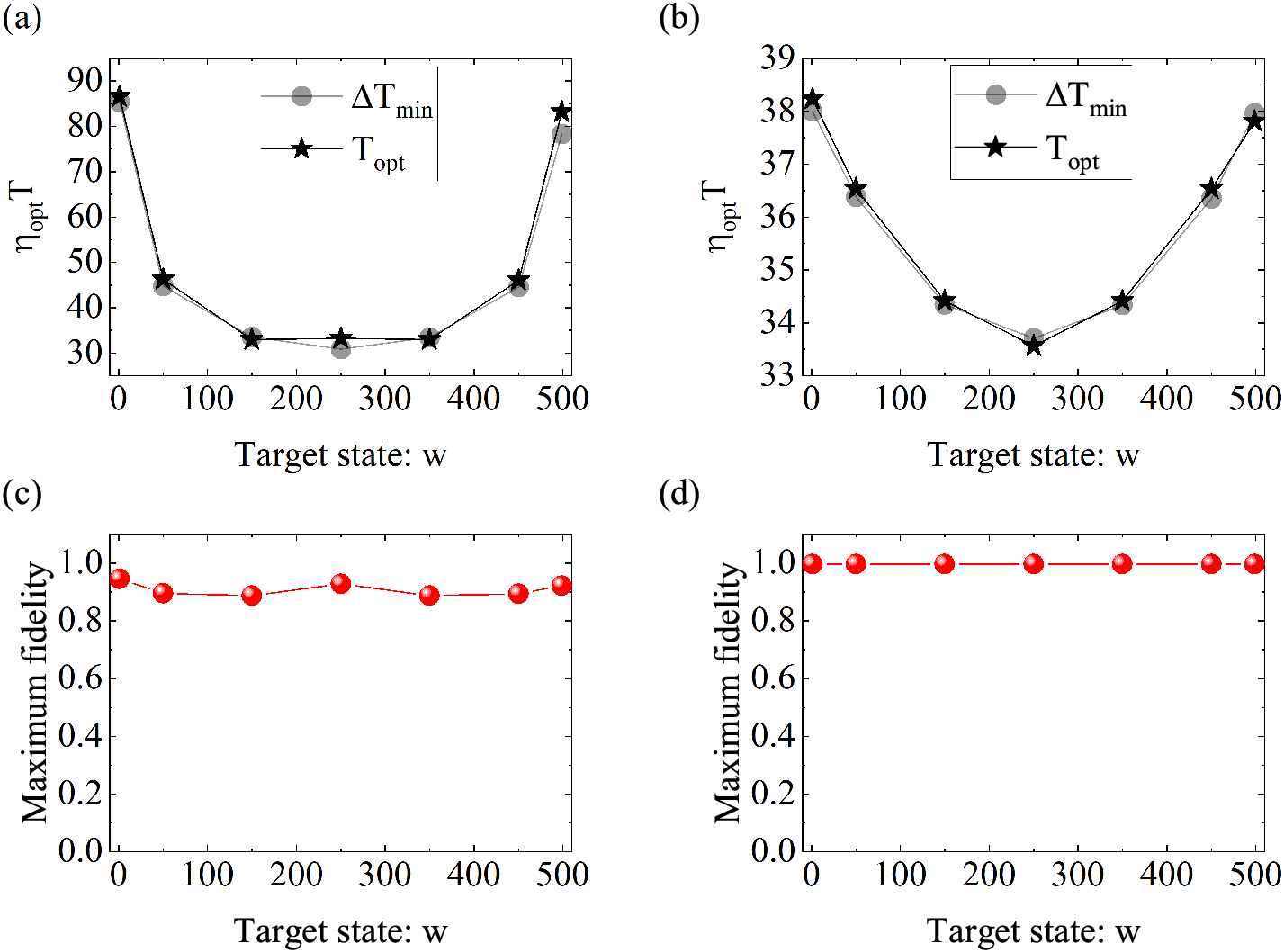}
\caption{(Color online)  Changes in search time and fidelity when selecting different target states, taking n=500 as an example, let target states $w=1$, 50, 150, 250, 350, 450, 499. (a, c) multi-power-law interactions in 1D optical lattices, and (b, d) exponentially decaying interactions in waveguide-QED.}
\label{4}
\end{figure}

\subsection{Boundary effects}

In the previous subsections, we assume that the position of the target is at $w=20$. For the complete graph such as in the cavity-QED system with infinite long-range interaction, the position of the target does not affect the search time. However, if the atom-atom interaction is not ideally infinite such as in the multi-power-law interaction and exponentially-decaying interaction cases, the connectivity of each node is not the same and thus the location of the marked node may affect the search time. For example, the atoms close to the boundaries (e.g., the leftmost and the rightmost) are the least interacting with the rest of the atoms, and the searching time for these nodes is expected to be longer. To demonstrate this boundary effect, we calculate the search time and fidelity when different locations of nodes are selected in the cases of multi-power-law interactions in a 1D optical lattice and exponentially decaying interactions in the waveguide-QED within the bandgap. Take n=500 as an example, we compare the results when $w=1$, 50, 150, 250, 350, 450, and 499, and the results are shown in Fig. 3. From Fig. 3(a) and 3(b), it is clearly seen that when the target node is close to the boundary, the search time is longer which is consistent with the expected prediction. The corresponding searching fidelities are also shown in Fig. 3(c) and 3(d) from which we can see that the fidelity does not significantly depend on the location of the marked node.

\section{Quantum search with noise effect}

\begin{figure*}[htbp]
\label{Fig3} \centering\includegraphics[width=17cm]{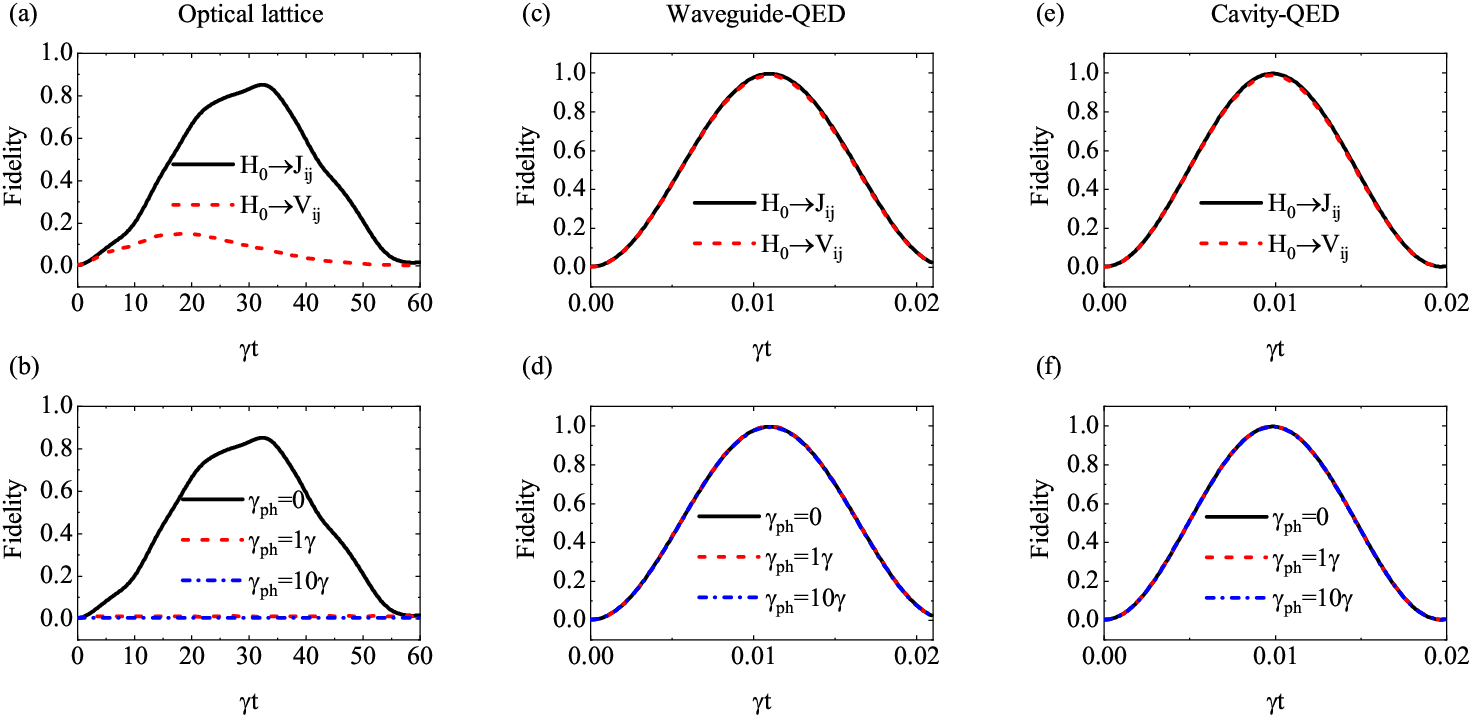}
\caption{(Color online)  Quantum search fidelity with dissipation effects: free-space optical lattice with mixture-power-law interaction with decay (a) and different dephasing rates (b). Waveguide-QED with exponentially decaying interactions ($\Gamma=20\gamma$ and $\kappa=0.001$) with collective decay (c) and different dephasing rates (d). Cavity-QED with ideal long-range interaction ($J_{ij}=10\gamma$) under decay (e) and different dephasing rates (f).  Number of atoms: $n=256$.}
\label{3}
\end{figure*}

In the previous section, we investigate the quantum search under three different physical systems and show that near quadratic speedup can be realized in all three cases without considering the dissipation effects. However, in real physical systems dissipations including decay and dephasing need to be taken into account. Here, we consider the maximum fidelity (i.e., success probability) of the quantum search under decay and dephasing using two methods. The first one is based on the master equation which is given by 
\begin{eqnarray}
\dot{\rho}_{s}(t)=-i[H_{coh},\rho_{s}(t)]+\mathcal{L}[\rho_{s}(t)],
\label{master}
\end{eqnarray}%
where $H_{coh}=\sum_{i<j}J_{ij}(\vert i \rangle \langle j \vert+\vert j \rangle \langle i \vert)+\eta H_{t}$ is the coherent part of the search Hamiltonian and the dissipation is captured by the Lindblad term $\mathcal{L}[\rho_{s}(t)]$. For the decay process, $\mathcal{L}[\rho_{s}(t)]=-\frac{1}{2}\sum_{ij}\Gamma_{ij}[\sigma_{i}^{+}\sigma_{j}^{-}\rho_{s}(t)+\rho_{s}(t)\sigma_{i}^{+}\sigma_{j}^{-}-2\sigma_{j}^{-}\rho_{s}(t)\sigma_{i}^{+}]$ and for the dephasing process $\mathcal{L}[\rho_{s}(t)]=-\frac{\gamma_{ph}}{2}\sum_{j}[\sigma_{j}^{z}\sigma_{j}^{z}\rho_{s}(t)+\rho_{s}(t)\sigma_{j}^{z}\sigma_{j}^{z}-2\sigma_{j}^{z}\rho_{s}(t)\sigma_{j}^{z}]$ with $\gamma_{dp}$ being the dephasing rate. The second method is by evolving the system under the effective search Hamiltonian $H_{s}=\sum_{ij}(J_{ij}+i\Gamma_{ij}/2))\vert i \rangle \langle j \vert+\eta H_{t}$ where the collective decay process has been included via the $i\Gamma_{ij}/2$ terms (see Table I).   For the dephasing process, we can model the system by adding random local field fluctuations to the diagonal elements of the system Hamiltonian. \textbf{These local fluctuations are not static but randomaly flucuate with time. For each time step of evolution, we randomly generate small fluctuation for each qubit, and these fluctuations are sampled from a Gaussian distribution whose mean is $0$ and standard deviation is set by the depahsing rate $\gamma_{ph}$ \cite{25Lewis2021}}.  By running the quantum search for many times (e.g., 500 times in our simulation), the results can be obtained by averaging over all the outputs. We compare the results using these two methods and find that they are consistent with each other (see Fig. 5 in the appendix). This indicates the validity of using the effective Hamiltonian. Since the method based on the master equation is very time-consuming, here we mainly use the method based on the effective Hamiltonian to evaluate the performance of our quantum search when $n>100$.

The results are shown in Fig. 4 where the first row compares the search fidelity with and without the decay process and the second row shows the search fidelity under different dephasing rates. \textbf{In Figs. 4(a), 4(c), and  4(e)}, the solid black line represents the case without decay, and the dotted red line represents the case considering decay. For the free-space optical lattice, the decay rate has a significant impact on the search fidelity, the maximum fidelity becomes very small after considering the decay (Fig. 4(a)). In contrast, the fidelities in the waveguide-QED system and the cavity-QED system with decay are almost the same as those without decay (Fig. 4(c) and 4(e)). The main reason is that the energy gaps in these two systems are very large due to the enhanced atom-atom interaction and the time required for the quantum search is very short which can significantly suppress the effect of dissipation. Similar phenomena also occur with dephasing.  In Fig. 4(b, d, f), we compare the search fidelity under three different dephasing rates ($\gamma_{ph}=0, 1\gamma$, and $10\gamma$). It is clearly seen that dephasing can also significantly reduce the searching fidelity in the free-space optical lattice system while it has little effect in the other two systems. Thus, both the atom chain dispersively coupled to the cavity and the waveguide-QED system within the bandgap are better candidate systems for realizing the optimal quantum search due to the enhanced long-range atom-atom interactions.

\begin{figure*}[htbp]
\label{Fig5} \centering\includegraphics[width=17cm]{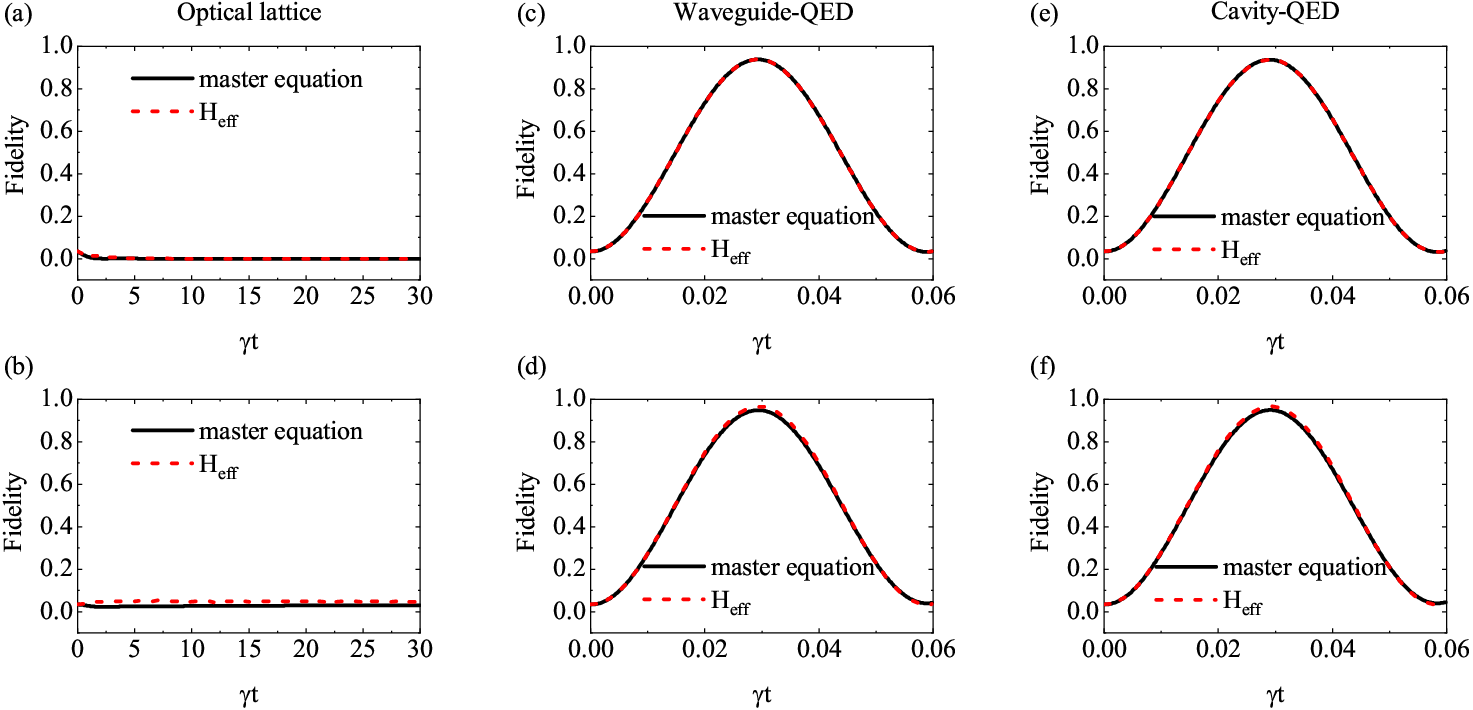}
\caption{(Color online) The search fidelity with dissipation effects calculated by master equation method and effective Hamiltonian method.  (a, c, e) with collective decay, (b, d, f) with pure dephasing $\gamma_{ph}=1\gamma$. (a, b) for optical lattice, (c, d) for waveguide-QED, and (e, f) for cavity-QED. Other parameters: n=30 and w=8. }
\label{5}
\end{figure*}

\section{SUMMARY}

We investigate the quantum spatial search problem in three different physical systems, i.e., atom array trapped in an optical lattice, atom array coupled to 1D waveguide within the bandgap, and atom arrays dispersively coupled to a good cavity. We find that close to quadratic speedup in quantum search can be achieved in all three systems when there are no dissipations. However, when there are dissipations including decay and dephasing, the waveguide-QED and cavity-QED system with enhanced collective interaction can still provide quadratic speedup with high success probability, while the success probability in the atom array trapped in an optical lattice is extremely low. These results indicate that the waveguide-QED and cavity-QED systems with long-range atom-atom interaction are better systems for demonstrating the optimal quantum search if the noise effect is considered. This is because the atom-atom interactions in the waveguide-QED and cavity-QED systems can be much larger than those in the free-space case even if two atoms are far apart in space. The enhanced collective long-range atom-atom interaction can result in a much larger spectra gap which can significantly reduce the search time and suppress the effects of dissipation.  Our results here can provide helpful instructions for realizing quantum search in real physical systems in the NISQ era.

\appendix

\section{Two methods for calculating the dissipation effects}

The dissipation effects including the decay and dephasing processes can be accounted for either by the Lindblad-form master equation or the effective Hamiltonian methods as illustrated in Sec. IV. Here, we compare the results of these two methods. 

Take the number of atoms to be $n=30$, and the target state is $w=8$ as an example. The results of search fidelity using either the master equation or the effective Hamiltonian methods are shown in Fig. 5 where the three subfigures in the first row present the results with decay while the three subfigures in the second row present the results with dephasing. It is clearly seen that the fidelities obtained by these two methods are almost the same for both the decay and dephasing processes. These results confirm the validity of the method using the effective Hamiltonian. Since the method calculated by the master equation is very time-consuming where we can only calculate the cases with tens of atoms. Thus, we mainly use the method based on effective Hamiltonian to calculate the performance of our search algorithm especially when the number of atoms is greater than 100.

\section*{Acknowledgments}

This work was supported by the National Key R\&D Program of China (Grant No. 2021YFA1400800), the Key Program of National Natural Science Foundation of China (Grant No. 12334017), the Key-Area Research and Development Program of Guangdong Province (Grant No.2018B030329001), the Guangdong Basic and Applied Basic Research Foundation (Grant No. 2023B1515040023), and the Natural Science Foundations of Guangdong (Grant No.2021A1515010039).

\end{document}